\documentclass[11pt,a4paper]{article}

\pdfoutput=1

\usepackage{jcappub}

\usepackage{epsfig}
\usepackage{dcolumn}
\usepackage{bm}
\usepackage{amssymb}
\usepackage{amsmath}
\usepackage{amsfonts}
\usepackage{graphicx}
\usepackage{psfrag}

\def\R{{\cal R} }

\subheader{\begin{flushright}
UTTG-06-10\\
TCC-015-10
\end{flushright}}

\title{Constraining the Inflationary Equation of State}

\author[a]{Lotty Ackerman,}
\author[a,b]{Willy Fischler,}
\author[a,b]{Sandipan Kundu}
\author[a,b]{and Navin Sivanandam}

\affiliation[a]{Texas Cosmology Center, University of Texas, Austin, TX 78712}
\affiliation[b]{Theory Group, Department of Physics, University of Texas, Austin, TX 78712}

\emailAdd{lotty@astro.as.utexas.edu}
\emailAdd{fischler@physics.utexas.edu}
\emailAdd{sandyk@physics.utexas.edu}
\emailAdd{navin.sivanandam@gmail.com}

\abstract{We explore possible constraints on the inflationary equation state: $p=w\rho$. While $w$ must be close to $-1$ for those modes that contribute to the observed power spectrum, for those modes currently out of experimental reach, the constraints on $w$ are much weaker, with only $w<-1/3$ as an a priori requirement. We find, however, that limits on the reheat temperature and the inflationary energy scale constrain $w$ further, though there is still ample parameter space for a vastly different (accelerating) equation of state between the end of quasi-de Sitter inflation and the beginning of the radiation-dominated era. In the event that such an epoch of acceleration could be observed, we review the consequences for the primordial power spectrum.}

\arxivnumber{1007.3511}

\begin{document}
\maketitle
\flushbottom

\section{Introduction}
The theory of inflation has the twin properties of explaining away (at least to some degree) many of the puzzles of the very early universe in a broad class of general models, whilst simultaneously remaining remarkably hard to pin down in specific detail\footnote{For an alternative solution to some of these puzzles see \cite{Willy}}. The key success of the inflationary theory is its prediction of an almost scale invariant power spectrum of primordial fluctuations \cite{Starobinsky:1982ee, Hawking:1982cz, Guth:1982ec, Bardeen:1983qw, Mukhanov:1985rz}, a prediction that is borne out by what we measure of the perturbations in the cosmological fluid today. However, the limited information we can extract about this power spectrum means that a wide class of inflationary models are viable. Moreover, the particular nature of that class of models and the constraints upon them are predicated on many assumptions about both the nature of inflation and the evolution history of the universe.

Whilst inflation is typically modeled by a (possibly multi-component) scalar field, the essential feature is that the expansion of universe undergoes a temporary accelerating period, where $\ddot{a}>0$. During accelerated expansion fluctuations ``exit'' the horizon (i.e. become larger than the physical size of the temporary event horizon). Then, once the expansion begins to decelerate, these fluctuations ``re-enter'' the horizon in reverse order from their exit. This means that the smallest scale perturbations we observe were produced last, closest to the start of decelerated expansion. 

\begin{figure}
\centering
\includegraphics[width=\textwidth]{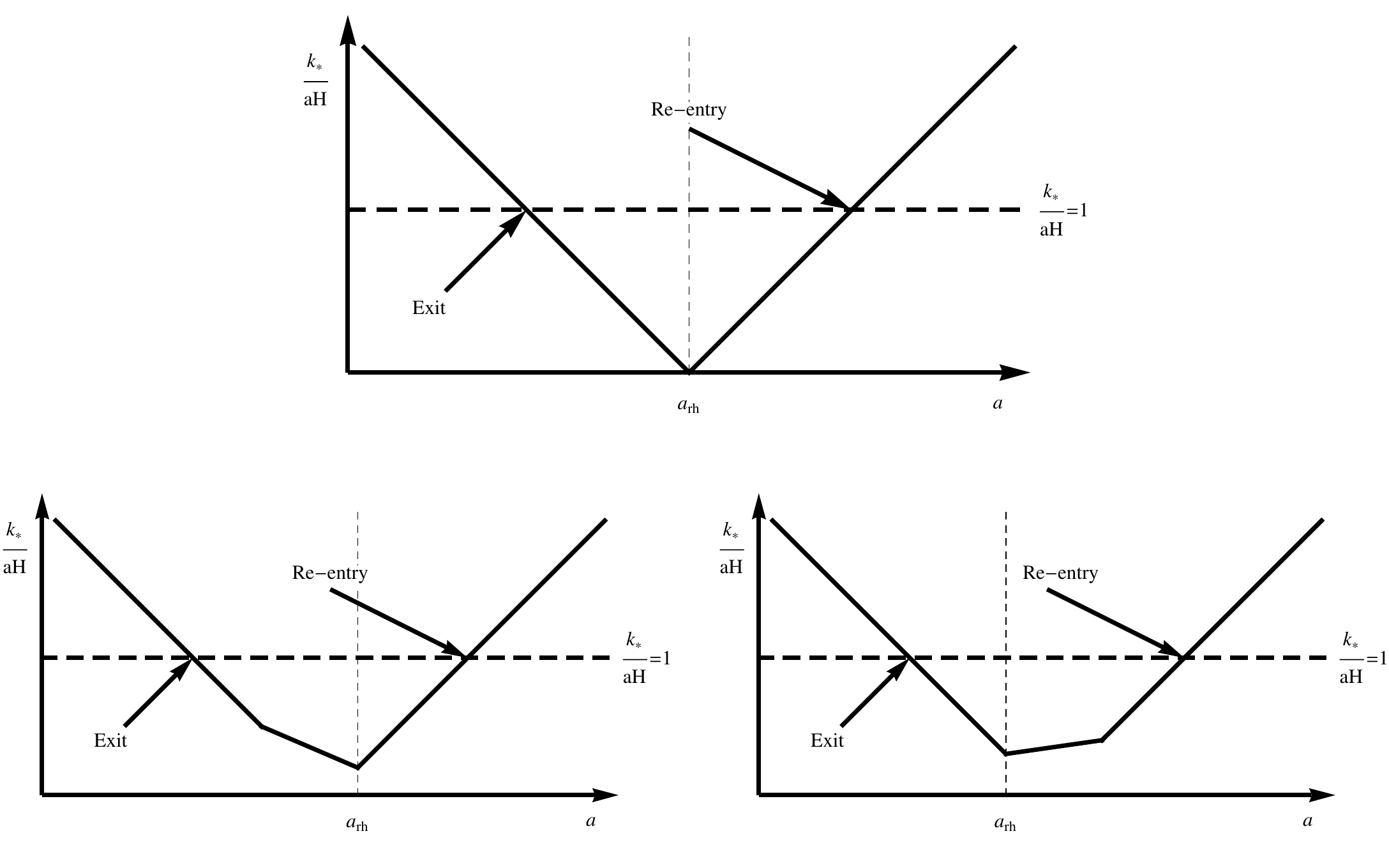}
\caption{\small{Schematic evolution of the quantity $k/aH$ for a particular scale, $k_*$. Changing the equation of state before(after) the end of inflation changes the negative(positive) part of the slope}}
\label{enterexit}
\end{figure}

The top graphic in figure \ref{enterexit} schematically illustrates this, with a logarithmic plot of the quantity $k_*/aH$. This quantity characterizes the size of the comoving horizon relative to a particular scale, $k_*$, with $k_*/aH=1$ corresponding to the scale exiting (or re-entering) the horizon. The slope of the line is related to the dynamics (or, in the language we will be using presently, the equation of state) of the cosmological fluid -- when the slope is negative(positive), the expansion is accelerating(decelerating).

In this paper we eschew the aforementioned scalar field paradigm, and rather consider the equation of state, bounding, as best we can, the equation of state parameter $w$, defined by $p=w\rho$. The Friedman equations (along with the null energy condition) mean that a positive $\ddot{a}$ implies that $-1<w<-1/3$. Changing the equation state changes the tilt of of the slope of the line in figure \ref{enterexit}. We are primarily interested in changes to the negative part of the slope (i.e. the accelerated expansion) that are consistant the current observational constraints. One could also consider bounds on the equation of state during decelerated expansion, which has been done by, for example, \cite{Boyle:2007zx} and \cite{Adshead:2010mc}. The two possibilities are schematically shown respectively on the left and right of the lower half of figure \ref{enterexit}. As we shall see, the broad cause of the uncertainty in the inflationary equation of state (and any uncertainity in the equation of state immediately after inflation) it that we only have a direct probe of the primordial spectrum for a subset of the e-foldings that correspond to our current observable universe.

An equation of state analysis of inflation has been carried out recently by Ilic, Kunz, Liddle and Frieman in \cite{Ilic:2010zp}, our approach is somewhat different, in particular we explore possible constraints on $w$ outside of those directly imposed by the power spectrum. Said power spectrum is almost scale-invariant from the size of the observable universe down to scales of around a Mpc; as the authors show in \cite{Ilic:2010zp}, this places tight bounds on the deviation of $w$ from $-1$. However, as we will describe below, $w$ is less tightly constrained for those periods of accelerated expansion corresponding to shorter length scales.

We explore the possibility of a significantly varying $w$ before the start of decelerating expansion by dividing the inflationary epoch into two distinct periods. The first, which for clarity we will refer to as ``inflation'', has $w\sim-1$ and is responsible for producing the nearly scale-invariant power spectrum we see on large scales. The second, from now on we call this ``accelerated expansion'', lasts as long as is necessary to be consistent with particular choices of the reheat scale (where deceleration begins at reheating) and the inflation-acceleration transition scale.

In section \ref{fluctuation} we review the derivation of the mode functions, and find an expression for $n_s$ in terms of $w$. In section \ref{observable}, we outline our model of the thermal history of the universe and discuss how constraints on observable wavenumbers in the power spectrum serve to bound $w$ (and thus $n_s$) during a period of accelerated expansion. We finish with some brief concluding remarks on what can be learned from this approach.

\section{Fluctuations During Accelerated Expansion}\label{fluctuation}
\subsection{Not Power-Law Inflation}
Before marching onwards with our calculations, we should note that periods of expansion with $w\neq-1$ (but $<-1/3$) give rise to what is  more commonly known as power-law inflation, where the scale factor grows like $t^p$ ($p>1$ is a function of $w$ -- the detailed solution can be found below). This is a well studied situation \cite{Abbott:1984fp, Lucchin:1984yf, Lyth:1991bc} in inflationary cosmology; it arises, for example, when we have a single scalar field with an exponential potential. We should note, however, that our setup is different in two important ways. Firstly, we are positing accelerating expansion in addition to inflation -- i.e. it is inflation that provides the nearly scale-invariant power spectrum we need to explain CMB and matter power spectrum data. All we allow our period of accelerating expansion to do is to provide additional e-foldings as needed to solve the horizon problem and to change (possibly) the power spectrum at the highest wavenumbers.

Furthermore, our treatment of the quantum vacuum for the modes produced during accelerated expansion is different. Rather than assuming a Bunch-Davies vacuum state \cite{Bunch:1978yq, Birrell:1982ix}, we instead take the modes during inflation as an initial condition on which to match those produced during accelerated expansion -- this is equivalent to assuming a sudden change in the equation of state from $w\sim-1$ to $-1<w<-1/3$. For obvious reasons (i.e. we match only the shortest wavelength modes) this gives the same answer as assuming that the Bunch-Davies vacuum is the appropriate one to use for power law accelerating expansion. In the case of power-law inflation for sufficiently large powers ($w$ sufficiently close to $-1$) one is close enough to de Sitter space such that the de Sitter invariant $\alpha$-vacua give an appropriate family of privileged states from which the Bunch-Davies vacuum is chosen. For larger values of $w$ (smaller powers), while one can construct the Bunch-Davies state, it no longer belongs to such a privileged invariant set, and thus is no longer an obvious choice for an appropriate vacuum.

Caveats and qualifications out the way, let us now move on to calculating the fluctuation spectrum.

\subsection{Mode Functions}
We are interested in the scalar fluctuations produced during an epoch of accelerated expansion (with the equation of state is given by $p = w\rho$ and $-1<w<-1/3$). Following standard treatments \cite{Bardeen:1980kt, Lyth:1984gv} we have the gauge invariant quantity, ${\cal R}$, which for a single scalar field  is conserved outside the horizon during inflation ($\R =\Psi+H\delta\phi/{\dot\phi}$). The Mukhanov variable is:
\begin{equation}
v\equiv z \R \ ,
\label{mukhanovvariable}
\end{equation}
with $z$ is defined in terms of the background scalar field as:
\begin{equation}
z=\frac{a \frac{d \phi}{dt}}{H} = \frac{a \dot{\phi}}{H}\ .
\end{equation}
In terms of the equation of state parameter, $z^2 = \frac{3}{8\pi G} a^2 (1 + w)$. The Fourier transform of $v$ satisfies the Mukhanov equation \cite{Mukhanov:1985rz, Mukhanov:1990me}:
\begin{equation}
\frac{d^2 v_k}{d \tau^2}+\left(k^2-\frac{1}{z}\frac{d^2 z}{d \tau^2}\right) v_k=0\ ,
\label{mukhanovequation}
\end{equation}
where $\tau $ is the conformal time.

Solving this equation allows us to find the mode functions during both inflation and our second period of accelerated expansion. This, in turn, will let us find an expression for the spectral index. We summarize the results below.

\subsubsection*{Solutions}
During inflation, with $w=w_{\textrm{inf}}\sim-1$ and $a=a_fe^{H_f(t-t_f)}$, the mode function is given by (in the Bunch-Davies vacuum):
\begin{equation}
v_k= \frac{e^{-ik\tau}}{\sqrt{2k}}\left(1-\frac{i}{k\tau}\right) \ .
\label{modefunctioninf}
\end{equation}
From equation \ref{mukhanovvariable}, this give $\R_k$ as:
\begin{equation}\label{Rinflation}
\R_k=\sqrt{\frac{8\pi G}{3(1+w_{\textrm{inf}})}}\frac{1}{a}\frac{e^{-ik\tau}}{\sqrt{2k}}\left(1-\frac{i}{k\tau}\right)
\end{equation}

During an epoch of accelerated expansion (matching the scale factor and energy density to the prior inflationary period) the scale factor evolves as:
\begin{equation}
a=a_f\left[\frac{3}{2}H_f (1+w)(t-t_f)+1\right]^{\frac{2}{3(1+w)}} \ ,
\label{a}
\end{equation}
with $H$ given by:
\begin{equation}
H=H_f\left(\frac{a_f}{a}\right)^{\frac{3}{2}(1+w)} \ ,
\label{h}
\end{equation}
and $\tau$ defined in terms of $a$ by:
\begin{equation}
\tau=\frac{1}{(3w+1)a_fH_f}\left[2\left(\frac{a}{a_f}\right)^{\frac{3w+1}{2}} -\quad 3(w+1)\right]
\end{equation}
This allows us to write equation \ref{mukhanovequation} as:
\begin{equation}
\frac{d^2 v_k}{d \tau^2}+\left[k^2- \frac{1-3w}{2 \tau_f^2\left(1-\frac{3w+1}{2}\frac{\tau-\tau_f}{\tau_f}\right)^2}\right] v_k=0\ ,
\label{epochequation}
\end{equation}
where $\tau_f=-\frac{1}{a_f H_f}$. The general form of the solution is given by a linear combination of Hankel functions:
\begin{equation}
v_k=A(k)\sqrt{x}H_n^{(1)}\left[\frac{kx}{Q}\right] + B(k) \sqrt{x} H_n^{(2)}\left[\frac{kx}{Q}\right]\ .
\label{epochsolution}
\end{equation}
With
\begin{equation}\label{Qnx}
Q=-\frac{3w+1}{2}> 0, \qquad n=\frac{3(1-w)}{4Q}, \qquad x=Q(\tau - \tau_f)+\tau_f\ .
\end{equation}
Then $\R$ during this period is given by:
\begin{equation}
\R_k= \sqrt{\frac{8\pi G}{3(1+w)}}\frac{1}{a}\left[A(k)\sqrt{x}H_n^{(1)}\left[\frac{kx}{Q}\right] + B(k) \sqrt{x} H_n^{(2)}\left[\frac{kx}{Q}\right]\right]
\end{equation}

The functions $A(k)$ and $B(k)$ can be found by matching this solution with that given in equation \ref{Rinflation} -- this, of course, assumes an instantaneous transition from inflation to a second period of acceleration. Doing this for all values of $k$ is non-trivial, in particular for $k$ corresponding to the horizon size, only numerical evaluation of $A(k)$ and $B(k)$ is possible. However, for modes asymptotic in $|k\tau_f|$ it is relatively straightforward to find the curvature perturbation. Far outside the horizon, $|k\tau_f|<<1$, the curvature perturbation is ``frozen-in'' and curvature perturbation remains at the constant value set by the inflationary epoch. Deep inside the horizon, $|k\tau_f|>>1$ matching solutions with equation \ref{Rinflation}, we obtain (for $\tau>\tau_f$):
\begin{equation}\label{Raccel}
\R_k=\frac{\pi}{2a}\sqrt{\frac{8 G}{3 Q (1+w_{\textrm{inf}})}} \frac{\left(1-\frac{i}{k\tau_f}\right)}{\left(1-\frac{i}{2}\frac{1+\frac{1}{Q}}{k\tau_f}\right)}\; \exp\left[-i\left(n \pi/2+ \pi/4- \frac{3k\tau_f}{2}\frac{1+w}{Q}\right)\right]\; \sqrt{x}\; H_n^{(2)}\left[\frac{kx}{Q}\right]\
\end{equation}

While it is relatively easy to extract observables (in particular the value of $n_s$) from the above expression (we do so below), we should note that if a secondary period on inflation produced visible consequences, the largest scales at which such effects could be observed do not correspond to
either of the two limits discussed above. Rather, since inflationary perturbations would be responsible for large scale perturbations that we already measure, from the size of the observable universe down to scales less than $\sim1\rm{Mpc}$, the largest scales from a secondary period of accelerated expansion would be those produced right after the end of inflation, when $k\tau_f\sim1$. At these scales, we have to match our solutions numerically. We have carried out numerical investigation of the matching for $k\tau_f\sim1$, and find that the form of $v_k$ approaches the asymptotic results within a few percent for $0.5\lesssim k\tau_f\lesssim2$. Accordingly, we concentrate on the form of the mode function given in equation \ref{Raccel} in order to obtain an analytic expression for the primordial power spectrum of a secondary period of acceleration, though we note that a more precise numerical calculation would be needed in the event of an actual measurement.

\subsubsection*{The Spectral Index $n_s$}
In order to construct a physical observable, we calculate the scalar spectral index. To do this we take the superhorizon limit, $|kx/Q|<<1$,\footnote{One can check this is indeed the superhorizon limit by finding expressions for both the physical wavenumber, $k/a$, and the event horizon, $a\int_a^{\infty}\frac{1}{a'^2H(a')} \mathrm{d}a'$, in terms of the quantities $Q$ and $x$ defined in equation \ref{Qnx}.} and use the asymptotic behavior of the Hankel function for  $|y|<<1$ (this result, of course, only applies to modes which also have $|k\tau_f|>>1$ -- i.e. modes which were well inside the horizon at the end of inflation):
\begin{equation}
H_{n}^{(1,2)}[y] \rightarrow \mp \frac{i \Gamma[n]}{\pi} \left( \frac{y}{2}\right)^{-n}\ .
\end{equation}
From this we obtain:
\begin{equation}
\R_k \rightarrow i \Gamma[n]\sqrt{\frac{8G}{3(1+w_{\textrm{inf}})}}\frac{Q^{n+\frac{1}{2}}}{1+Q}\left(k^{-n}a^{-1}x^{(\frac{1}{2}-n)}\right)\exp\left[-i\left(n \pi/2+ \pi/4- \frac{3k\tau}{2}\frac{1+w}{Q}\right)\right]\ .
\label{finalmodefunction}
\end{equation}
Therefore, $|\R_k|^2 \propto k^{-2n}$, and the scalar spectral index, $n_s$ (defined through $|\R_k|^2 \propto k^{-4+n_s}$), is given by:
\begin{equation}\label{nsw}
n_s - 1 = 3-2n=\frac{6(1+w)}{(1+3w)}\ .
\end{equation}
Clearly, for $-1<w<-1/3$, we have $1>n_s>-\infty$. Having found the relationship between $w$ and $n_s$ our next goal is to see how constraints can be brought to bear on these quantities. We do this by considering the thermal history of the universe.

\section{Constraints From Thermal History}\label{observable}
As discussed above, fluctuations produced during a period of accelerated expansion exit the horizon, and when they do so the associated curvature perturbation does not evolve (assuming that a single component fluid is driving the acceleration), until it re-enters the horizon at some later time, when the expansion is decelerating. Since our posited period of additional accelerated expansion takes place after inflation, it will affect the power spectrum at shorter scales. By considering the evolution history of the universe and the various bounds observations of the power spectrum place on said history, we can constrain some of the parameters of any additional period of accelerated expansion.

We assume a sequence of epochs after inflation, with a variety of different equations of state: during inflation we have $w=-1$, during our accelerated period of expansion we have $-1<w_{a}<-1/3$, then from $t=t_{rh}$ (the time of reheating) the universe evolves in the conventional manner as an appropriate mixture of radiation, matter and $\Lambda$ (this, of course, assumes an instantaneous reheat). Solving the Friedmann equations in the various different epochs and matching the scale factor and energy density across boundaries gives the solutions shown in table \ref{equations} (for clarity we've glossed over the details of scale factor evolution after reheating, but this will be taken into account below when finding constraints).

\begin{table}[h]
\centering
\begin{tabular}{c|c|r c l}
$t$ & $w$ & \multicolumn{3}{c}{Solutions}\\
\hline
$t<t_f$ & $\sim-1$ & $a$ & $=$ & $a_fe^{H_f\left(t-t_f\right)}$\\
& & $\rho$ & $=$ & $\rho_f$\\
& & $H$ & $=$ & $\sqrt{\frac{8\pi G}{3}\rho_f}$\\
\hline
$t_f<t<t_{rh}$ & $-1<w_{a}<-1/3$ & $a$ & $=$ & $a_f\left[\frac{3}{2}H_f\left(1+w_{a}\right)\left(t-t_f\right)+1\right]^{\frac{2}{3(1+w_{a})}}$\\
& & $\rho$ &  $=$ & $\rho_f\left(\frac{a_f}{a}\right)^{3(1+w_{a})}$\\
& & $H$ & $=$ & $H_f\left(\frac{a_f}{a}\right)^{\frac{3}{2}(1+w_{a})}$\\
\hline
$t>t_{rh}$ & $1/3$ & $a$ & $=$ & $a_{rh}\left[2H_{rh}\left(t-t_{rh}\right)+1\right]^{\frac{1}{2}}$\\
& & $\rho$ & $=$ & $\rho_{rh}\left(\frac{a_{rh}}{a}\right)^4$\\
& & $H$ & $=$ & $H_{rh}\left(\frac{a_{rh}}{a}\right)^2$
\end{tabular}
\caption{Cosmological solutions for epochs of interest. These equations are somewhat over-parameterized, all parameters can be expressed in terms of $\rho_f$ and various scale factor normalizations.}
\label{equations}
\end{table}

We can use the above history to give us a bound on the combined number of e-foldings from inflation and our accelerated period of expansion. Noting that during radiation-domination and other decelerating periods of expansion the particle horizon grows faster than the universe (and thus faster than the perturbations in the cosmological fluid), we see that in the far past modes that are currently inside the horizon were far outside it. This is the horizon problem, and it is solved by having a period of expansion where the horizon grows slower than the universe, which lasts long enough to ensure that scales within horizon today were also within the horizon in the early universe. The number of e-foldings that are required to solve the horizon problem is fixed by the amount of horizon growth relative to the size of the universe since deceleration began \cite{Weinberg:2008zzc}, or:
\begin{equation}
N_0=\ln\left[\frac{a_{rh}H_{rh}}{a_0H_0}\right]=\ln\left[\frac{\rho_{rh}^{1/4}}{0.037h~\rm{eV}}\right] \ .
\label{efoldings}
\end{equation}
$h$ is the Hubble constant in units of $100\rm{km}~\rm{s}^{-1}~\rm{Mpc}^{-1}$. Furthermore, the number of e-foldings that took place between the time at which the mode corresponding to the size of the horizon today previously held that honor, and the time of reheating is given by:
\begin{equation}
N_{obs}=N_0+\ln\left[\frac{H[t_{q_0}]}{H[t_{rh}]}\right] \ .
\end{equation}
$t_{q_0}$ is the time at which the mode corresponding to the size of the horizon today was produced. Since for our model this mode was produced during inflation, and since $H$ is approximately constant during the inflationary epoch, we can replace $t_{q_0}$ with $t_f$:
\begin{equation}\label{nobsgen}
\ln\left[\frac{H[t_f]}{H[t_{rh}]}\right]_{max}=\frac{3}{2}\left(1+w_a\right)\ln\left[\frac{a_{rh}}{a_f}\right]=\frac{3}{2}\left(1+w_a\right)N_a \ .
\end{equation}
$N_a$ is the number of e-foldings from accelerated expansion. Thus we have:
\begin{equation}\label{nmax}
N_{obs}=N_0+\frac{3}{2}\left(1+w_a\right)N_a \ .
\end{equation}

Incidentally, since $N_{obs}\geq N_0$ and since the observed scale-invariant spectrum of perturbations is at scales from the current size of the observable universe down, requiring that the primordial power spectrum (at large scales) is explained by inflation guarantees that the horizon problem will be solved. It's clear from equation \ref{efoldings} that the minimum required number of e-foldings is fewer for lower scale reheating. We cannot, however, push this scale arbitrarily low -- we know, for example that the universe was radiation dominated at the time of big bang nucleosynthesis, which corresponds to a scale of around $1~\rm{MeV}$ (in \cite{Hannestad:2004px} Hannestad finds $T_{rh}>4~\rm{MeV}$ or $>1~\rm{MeV}$ if reheating direct to neutrinos). We thus work with a conservative lower bound on the reheat scale of $\rho_{rh}^{1/4}\sim 10~\rm{MeV}$ or -- this would give $\gtrsim 19$ visible e-foldings (the precise bound depends also on the value of $N_i$)\footnote{One might complain that baryogenesis places tighter bounds on $T_{rh}$; while this is true for most models of baryogenesis, it can be avoided in the Affleck-Dine scenario \cite{Affleck:1984fy}.}. This analysis, and that which follows below, assumes instantaneous reheating -- if this were not the case the true value of $N_a$ would be somewhat smaller, with the rest of parameters adjusted accordingly.

In our scenario the e-foldings within the observable universe are divided between inflation and accelerated expansion. We can place a lower limit on the portion of visible e-foldings coming from inflation by considering the form of the power spectrum, in particular the size of the deviation from scale invariance. To do this we use a recent reconstruction of the power spectrum from Peiris and Verde \cite{Peiris:2009wp}. The authors reconstruct $n_s(q)$ for $0.0001\leq q~[h/\rm{Mpc}]\leq 3$, finding that $0.7\lesssim n_s(q)\lesssim 1.3$ -- the values of this bound depend on the details of the reconstruction, and it is somewhat tighter for $q$ away from the upper and lower limits. Let us define $q_{max}$ as the physical wavenumber corresponding to the smallest scale perturbation produced by inflation and $q_0$ as the physical wavenumber corresponding to the horizon today, then the number of visible e-foldings produced by inflation is given by:
\begin{equation}
N_i=\ln\left[\frac{q_{max}}{q_0}\right] \ .
\end{equation}
Then from equation \ref{nmax} and noting $N_{obs}=N_a+N_i$, we have:
\begin{align}\label{Naw}
N_a = & -\frac{2}{3w_a+1}\left(N_0 - N_i\right) \nonumber \\
= & -\frac{2}{3w_a+1}\left(\ln\left[\frac{a_{rh}H_{rh}}{a_0H_0}\right]-\ln\left[\frac{q_{max}}{q_0}\right]\right) \nonumber \\
= & -\frac{2}{3w_a+1}\ln\left[\frac{\rho_{rh}^{1/4}}{0.037h~\rm{eV}}\frac{q_0}{q_{max}}\right] \ .
\end{align}
$N_a$ can also be expressed in terms of the ratio of the scale factors at the beginning and end of accelerated expansion:
\begin{align}\label{Naw2}
N_a = & \ln\left[\frac{a_{rh}}{a_f}\right] \nonumber \\
= & \frac{1}{3(1+w_a)}\ln\left[\frac{\rho_f}{\rho_{rh}}\right] \ .
\end{align}
Bounding $\rho_f^{1/4}\lesssim10^{16}~\rm{GeV}$ (this bound comes from the amplitude of CMB fluctuations, $10^{-5}$ and the tilt-imposed upper limit of the slow roll parameter, $\epsilon\lesssim0.05$, see e.g. \cite{Weinberg:2008zzc} for details) we can use equations \ref{Naw} and \ref{Naw2} to find allowed values of $N_a$ and $w_a$ for different choices of $\rho_{rh}$.

\begin{figure}[!t]
\centering
\includegraphics[width=\textwidth]{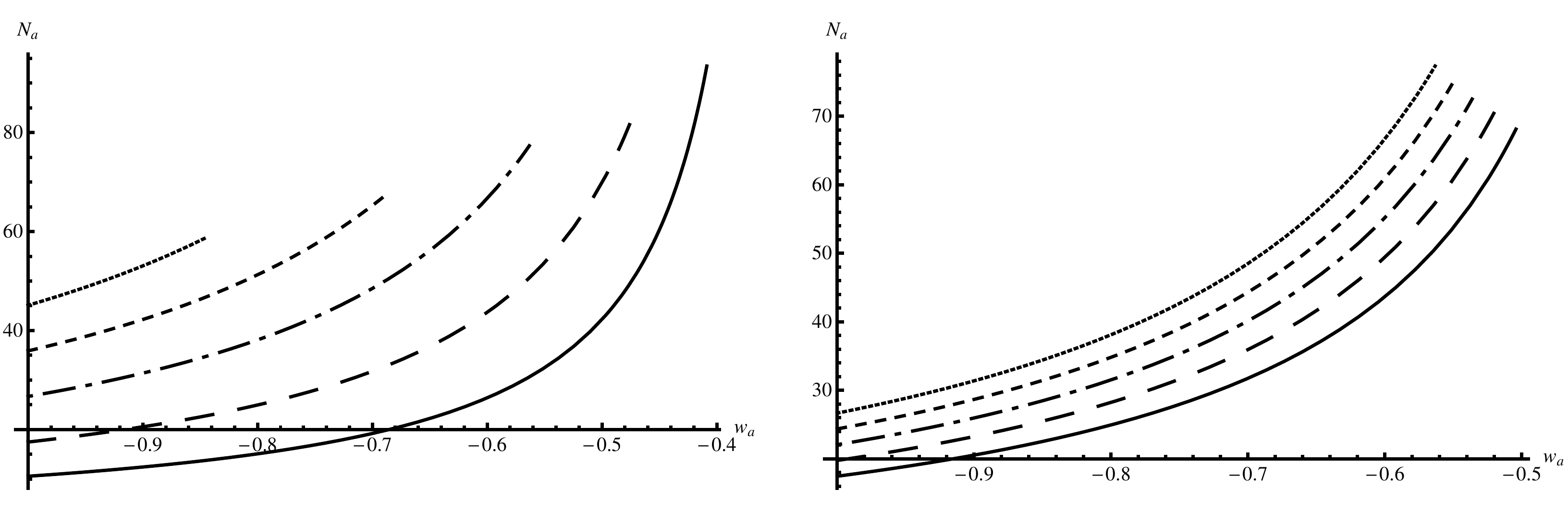}
\caption{\small{The left-hand plot shows the relationship between $N_a$ and $w$ for different choices of $\rho_{rh}^{1/4}$ (with $N_i\sim\ln10^{4}$): $10^{16}~\rm{MeV}$ for the uppermost dotted line, $10^{12}~\rm{MeV}$ for the dashed line, $10^{8}~\rm{MeV}$ for the dot-dash line, $10^{4}~\rm{MeV}$ for the widely spaced dashed line and $10~\rm{MeV}$ for the solid line. The right-hand plot shows the same for differing $N_i$ (with $\rho_{rh}^{1/4}\sim10^{8}~\rm{MeV}$): $\ln10^{4}$ for the uppermost dotted line, $\ln10^5$ for the dashed line, $\ln10^6$ for the dot-dash line, $\ln10^7$ for the widely spaced dashed line and $\ln10^8$ for the solid line.}}
\label{wNaplot}
\end{figure}

In figure \ref{wNaplot} we illustrate the nature of the $N_a$-$w_a$ relationship as we vary $\rho_{rh}^{1/4}$ (left-hand plot) and $N_i$ (right-hand plot); note that $\rho_f$ varies as one moves along the curves, each of which terminates when the upper limit for $\rho_f$ is reached. It is clear that for a sufficiently low reheat temperature our current observations of the power spectrum could be consistent with having a period of quasi de Sitter inflation followed by a period of accelerated expansion with an equation of state significantly different from $w\sim-1$. Moreover, such a situation can be arranged with $N_i$ sufficiently large such that the perturbations sourced by accelerated expansion are at length scales much smaller than those currently probed by experiment. On the other hand, if we wish to push up as close as possible to the current measurement of $N_i\sim\ln10^4$ and thus be on the boundary of having an observable signature of accelerated expansion, we find the most dramatic result we can have (i.e. the largest $w$), comes from having a minimal reheat temperature, and would give us $\sim93$ e-foldings of accelerated expansion with $w\sim-0.41$ and $n_s\sim-14.7$.

As our ability to measure the power spectrum on the shortest possible length scales improves, it is possible that we may be able to see the effects of a variation in the equation of state. However, given the difficulty of extracting information about the primordial power spectrum from scales much smaller than the scale at which local gravitational effects (i.e. not cosmology) dominates the physics it seems unlikely that any distinct signature of a second period of acceleration, unless its effects start at scales close to those at which we are currently confident of a scale-invariant spectrum. This restriction applies not only to the tilt, but also to other potential probes of the equation of state that rely on direct knowledge of the the spectrum (e.g. the running of $n_s$, higher order correlations etc.).

Although direct measurement of the spectrum is progressively harder at shorter and shorter distance scales, there are indirect probes that could be useful bounding the amplitude of primordial fluctuations. In particular, we know there must have been sufficient power at short distances to give rise to star production at an early enough times to produce the reionizened universe that we observe for $z\lesssim6$ (some of the relevant issues are discussed in \cite{Fernandez:2010gr}). In principle this requirement may conflict with the most extreme scenarios we consider above, since a strongly red-tilted spectrum means that considerably less power is present at short scales when compared to the $\delta\rho/\rho\sim10^{-5}$ that we have at large scales. However, although such constraints depend on the details of reionization in a manner beyond the scope of this paper, they seem unlikely to provide much stronger bounds on $N_a$ and $w_a$ for any given $q_{max}/q_0$ and $\rho^{1/4}_{rh}$.

Before moving on to our conclusions, let's consider what might happen if there was a secondary period of acceleration with a more complicated equation of state than that given by a constant $w$. To understand how such a period (distinguishable for inflation in principle) can be constrained in general it is helpful to consider again the relationships $N_a=N_{obs}-N_i$ and $N_{obs}\geq N_0$ (the latter is implied by equations \ref{nobsgen}). Together these give:
\begin{align}\label{Nabound1}
N_a \geq & N_0 - N_i \nonumber \\
= & \ln\left[\frac{\rho_{rh}^{1/4}}{0.037h~\rm{eV}}\right]-N_i \ .
\end{align}
Given the lower bounds on the reheat temperature and the number of e-foldings of almost scale-invariance it's clearly possible to have at least a few e-foldings of arising from a second accelerating epoch (with $O(10)$ MeV reheating and scale invariance down to a Mpc, there are 8 unobserved e-foldings). Higher reheating scales would leave more possible e-foldings hidden. The precise nature of the accelerating expansion that is responsible for those e-foldings outside our current observations is almost free of constraints, subject only to the requirement that the energy density must change by a sufficient amount between the end of inflation and the start of decelerated expansion after reheating. This leaves a large set of possibilities available for an alternative acceleration mechanism between the end of inflation and the start radiation-domination, though one would expect that unless the equation of state parameter is changing rapidly compared to the Hubble constant, the space of possibilities would be constrained in a similar manner to the constant $w$ case discussed here\footnote{In fact if one defines the parameter $w_{eff}=\frac{\int_1^{N_a} w}{N_a}$, one replace $w$ with $w_{eff}$ in the above expressions, to give the same restrictions, but now on $N_a$ and $w_{eff}$}.

\section{Conclusions}
In this paper we have explored the possibilities of a secondary period of accelerated expansion in the universe's history. With such a period immediately following inflation we show that the bounds imposed by considering current measurements of the power spectrum are relatively weak -- with reasonable reheat temperatures and inflationary scales it is relatively easy to fit in a secondary period of accelerated expansion with a vastly different equation of state to inflation and still have a nearly scale-invariant spectrum at scales from the size of the universe down to a Mpc and below.

\section*{Acknowledgements}
We would like to thank Eiichiro Komatsu and Paul Shapiro for helpful conversations. NS would like to thank Mustafa Amin for the same. This material is based upon work supported by the National Science Foundation under Grant No. PHY-0455649 and by the Texas Cosmology Center, which is supported by the College of Natural Sciences and the Department of Astronomy at the University of Texas at Austin and the McDonald Observatory.


\begin{thebibliography}{999}


\bibitem{Willy}
  T.~Banks and W.~Fischler,
  ``The holographic approach to cosmology,''
  arXiv:hep-th/0412097.
  \\
  T.~Banks and W.~Fischler,
  ``Holographic cosmology,''
  arXiv:hep-th/0405200.
  \\
  T.~Banks and W.~Fischler,
  ``Holographic cosmology 3.0,''
  Phys.\ Scripta {\bf T117}, 56 (2005)
  [arXiv:hep-th/0310288].
  \\
  T.~Banks and W.~Fischler,
  ``An holographic cosmology,''
  arXiv:hep-th/0111142.
  \\
  T.~Banks, W.~Fischler and L.~Mannelli,
  ``Microscopic quantum mechanics of the p = rho universe,''
  Phys.\ Rev.\  D {\bf 71}, 123514 (2005)
  [arXiv:hep-th/0408076].
  \\
  T.~Banks and W.~Fischler,
  ``M-theory observables for cosmological space-times,''
  arXiv:hep-th/0102077.

\bibitem{Starobinsky:1982ee}
  A.~A.~Starobinsky,
  ``Dynamics Of Phase Transition In The New Inflationary Universe Scenario And
  Generation Of Perturbations,''
  Phys.\ Lett.\  B {\bf 117}, 175 (1982).

\bibitem{Hawking:1982cz}
  S.~W.~Hawking,
  ``The Development Of Irregularities In A Single Bubble Inflationary
  Universe,''
  Phys.\ Lett.\  B {\bf 115}, 295 (1982).

\bibitem{Guth:1982ec}
  A.~H.~Guth and S.~Y.~Pi,
  ``Fluctuations In The New Inflationary Universe,''
  Phys.\ Rev.\ Lett.\  {\bf 49}, 1110 (1982).

\bibitem{Bardeen:1983qw}
  J.~M.~Bardeen, P.~J.~Steinhardt and M.~S.~Turner,
  ``Spontaneous Creation Of Almost Scale - Free Density Perturbations In An
  Inflationary Universe,''
  Phys.\ Rev.\  D {\bf 28}, 679 (1983).

\bibitem{Mukhanov:1985rz}
  V.~F.~Mukhanov,
  ``Gravitational Instability Of The Universe Filled With A Scalar Field,''
  JETP Lett.\  {\bf 41}, 493 (1985)
  [Pisma Zh.\ Eksp.\ Teor.\ Fiz.\  {\bf 41}, 402 (1985)].

\bibitem{Boyle:2007zx}
  L.~A.~Boyle and A.~Buonanno,
  Phys.\ Rev.\  D {\bf 78}, 043531 (2008)
  [arXiv:0708.2279 [astro-ph]].

\bibitem{Adshead:2010mc}
  P.~Adshead, R.~Easther, J.~Pritchard and A.~Loeb,
  JCAP {\bf 1102}, 021 (2011)
  [arXiv:1007.3748 [astro-ph.CO]].

\bibitem{Ilic:2010zp}
  S.~Ilic, M.~Kunz, A.~R.~Liddle and J.~A.~Frieman,
  ``A dark energy view of inflation,''
  arXiv:1002.4196 [astro-ph.CO].

\bibitem{Abbott:1984fp}
  L.~F.~Abbott and M.~B.~Wise,
  ``Constraints On Generalized Inflationary Cosmologies,''
  Nucl.\ Phys.\  B {\bf 244}, 541 (1984).

\bibitem{Lucchin:1984yf}
  F.~Lucchin and S.~Matarrese,
  ``Power Law Inflation,''
  Phys.\ Rev.\  D {\bf 32}, 1316 (1985).

\bibitem{Lyth:1991bc}
  D.~H.~Lyth and E.~D.~Stewart,
  ``The Curvature perturbation in power law (e.g. extended) inflation,''
  Phys.\ Lett.\  B {\bf 274}, 168 (1992).

\bibitem{Bunch:1978yq}
  T.~S.~Bunch and P.~C.~W.~Davies,
  ``Quantum Field Theory In De Sitter Space: Renormalization By Point
  Splitting,''
  Proc.\ Roy.\ Soc.\ Lond.\  A {\bf 360}, 117 (1978).

\bibitem{Birrell:1982ix}
  N.~D.~Birrell and P.~C.~W.~Davies,
  ``Quantum Fields In Curved Space,''
{\it  Cambridge, UK: Camb. Univ. Pr. ( 1982) 340p}

\bibitem{Bardeen:1980kt}
  J.~M.~Bardeen,
  ``Gauge Invariant Cosmological Perturbations,''
  Phys.\ Rev.\  D {\bf 22}, 1882 (1980).

\bibitem{Lyth:1984gv}
  D.~H.~Lyth,
  ``Large Scale Energy Density Perturbations And Inflation,''
  Phys.\ Rev.\  D {\bf 31}, 1792 (1985).

\bibitem{Mukhanov:1990me}
  V.~F.~Mukhanov, H.~A.~Feldman and R.~H.~Brandenberger,
  ``Theory of cosmological perturbations. Part 1. Classical perturbations. Part
  2. Quantum theory of perturbations. Part 3. Extensions,''
  Phys.\ Rept.\  {\bf 215}, 203 (1992).

\bibitem{Weinberg:2008zzc}
  S.~Weinberg,
  ``Cosmology,''
{\it  Oxford, UK: Oxford Univ. Pr. (2008) 593 p}

\bibitem{Affleck:1984fy}
  I.~Affleck and M.~Dine,
  ``A New Mechanism For Baryogenesis,''
  Nucl.\ Phys.\  B {\bf 249}, 361 (1985).

\bibitem{Hannestad:2004px}
 S.~Hannestad,
 ``What is the lowest possible reheating temperature?,''
 Phys.\ Rev.\  D {\bf 70}, 043506 (2004)
 [arXiv:astro-ph/0403291].

\bibitem{Peiris:2009wp}
  H.~V.~Peiris and L.~Verde,
  ``The Shape of the Primordial Power Spectrum: A Last Stand Before Planck,''
  Phys.\ Rev.\ D {\bf 81}, 021302 (2010) [arXiv:astro-ph/0912.0268].

\bibitem{Fernandez:2010gr}
  E.~R.~Fernandez and J.~M.~Shull,
  ``The Effect of Galactic Properties on the Escape Fraction of Ionizing
  Photons,''
  arXiv:1006.3519 [astro-ph.CO].

\end{thebibliography}
\end{document}